# Evaluating openEHR for storing computable representations of electronic health record phenotyping algorithms

Václav Papež[†], Spiros Denaxas[†], Harry Hemingway
Institute of Health Informatics, University College London
Farr Institute of Health Informatics Research, London UK
{v.papez, s.denaxas, h.hemingway}@ucl.ac.uk

*Abstract*— **Electronic Health Records (EHR) are data generated during routine clinical care. EHR offer researchers unprecedented phenotypic breadth and depth and have the potential to accelerate the pace of precision medicine at scale. A main EHR use-case is creating phenotyping algorithms to define disease status, onset and severity. Currently, no common machine-readable standard exists for defining phenotyping algorithms which often are stored in human-readable formats. As a result, the translation of algorithms to implementation code is challenging and sharing across the scientific community is problematic. In this paper, we evaluate openEHR, a formal EHR data specification, for computable representations of EHR phenotyping algorithms.**

*Keywords-electronic health records; phenotyping; standards*

## I. INTRODUCTION

Electronic Health Records (EHR) are structured, semi-structured and unstructured data generated during interactions of patients with healthcare and are increasingly utilized for research [1]. High-throughput genotyping and increased availability of EHR data are giving scientists the unprecedented opportunity to exploit routinely generated clinical data to deliver personalized interventions. EHR data can fundamentally alter the manner in which genetic association studies are performed and enable scientists to examine the association of genetic variants and traits in larger sample sizes and phenotypic breadth [2].

### A. EHR Phenotyping

A primary use-case of EHR data is *disease phenotyping:* the creation of computational algorithms that identify patients that have been diagnosed with a particular condition (e.g. acute myocardial infarction) and where applicable the disease onset and severity [3] using clinical information such as diagnoses, laboratory tests, symptoms, clinical examination findings, prescriptions, referrals and other data elements stored in EHR. The process of defining and validating EHR phenotypes poses significant challenges [4]. EHR data are not primarily collected for research and thus offer an indirect representation of the true patient state skewed by underlying healthcare processes (e.g. clinical guidelines, information systems, data standards). Challenges are amplified by the lack of a common standard for defining EHR-driven phenotypes, making their sharing and cross-site evaluation problematic. While phenotype components are structured and often annotated by controlled clinical terminologies, phenotype definitions, and their underlying algorithmic logic are expressed as free-text and/or graphical form which is not machine-readable. The translation from narrative to code (e.g. Structured Query Language) is problematic due to potential ambiguities.

EHR phenotyping algorithms should ideally be stored in a computable, standards-driven format to facilitate their systematic creation, sharing and re-use. This can be potentially enabled by formal EHR data specifications, such as openEHR (http://openehr.org. In this work, we evaluate the use of openEHR archetypes for storing deterministic EHR-derived phenotyping algorithms. We used diabetes as a case study since it exemplifies many of the associated challenges but our findings are generalizable to other diseases and syndromes.

---
[†] Both authors contributed equally to this work.

## II. BACKGROUND AND METHODS

### A. CALIBER

We used a deterministic diabetes phenotyping algorithm previously developed and validated in CALIBER [5]. CALIBER is a translational EHR research resource which links national, structured primary care, hospital care, disease registry, mortality data and socioeconomic information in the UK for ~10m patients. Primary care data are provided by the Clinical Practice Research Datalink, an anonymized national cohort of longitudinal data for all individuals registered with a general practitioner. Secondary care data are recorded in Hospital Episode Statistics (HES), a national database of administrative data used for hospital reimbursement. Finally, mortality and socioeconomic data are collected and curated by the Office of National Statistics (ONS).

Primary care data include diagnoses, referrals, symptoms, laboratory tests and clinical examination findings recorded using the Read controlled clinical terminology (a subset of SNOMED-CT). Medication prescriptions are organized using the British National Formulary, a structured resource for classifying all therapeutic agents prescribed in UK healthcare. Hospital care data include ranked diagnoses recorded using the International Classification of Diseases $10^{th}$ revision (ICD-10) terms and interventional procedures recorded using the OPCS Classification of Interventions and Procedures version 4 (OPCS4) terms. Mortality data are recorded using ICD-9 and ICD-10 and include the underlying cause of death and up to 15 contributory causes of mortality.

### B. Diabetes phenotyping algorithm

We used a previously validated [6] EHR phenotyping algorithm for identifying and classifying patients with diabetes into four non-overlapping groups: 1) patients with type 1 diabetes, 2) patients with type 2 diabetes, 3) patients with unspecified diabetes and 4) patients that are not diabetic.

The algorithm (Fig. 1) combines clinical information from specific diagnostic codes for type 1 and type 2 diabetes with less specific codes for 'insulin dependent diabetes' and 'non-insulin dependent diabetes' recorded across both primary and secondary care. Patients with diagnoses of both type 1 and type 2 diabetes are classified as patients with diabetes of unspecified type. The algorithm was designed to primarily identify patients with type 2 diabetes that can be on a variety of medications so it does not make use of medication data explicitly.

Phenotype components are stratified and named by datasource: *dm_gprd* represents a diagnosis of diabetes in primary care, *dm_hes* a diagnosis in secondary care. Within each component, diagnostic codes have been grouped by clinicians in terms of certainty (e.g. *historical, possible, confirmed*).

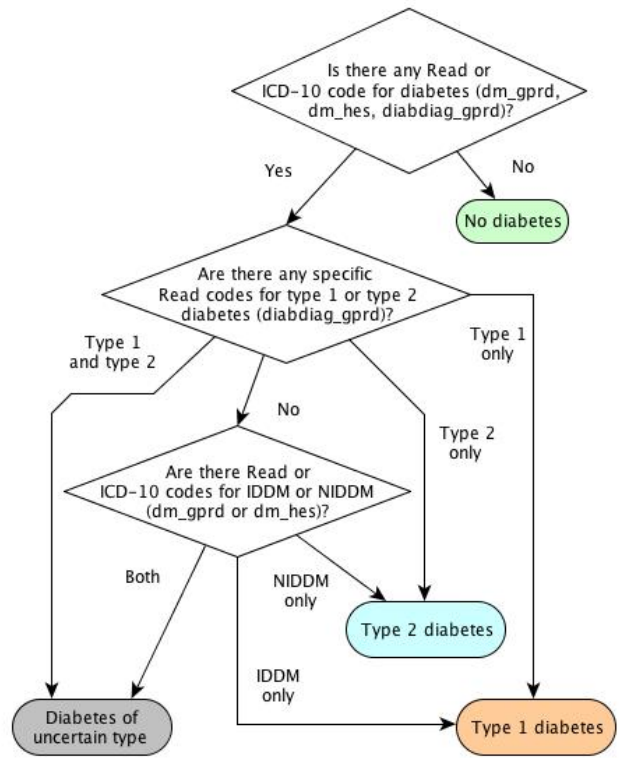

Figure 1. Diabetes EHR phenotyping algorithm classifies patients into four groups: type 1 diabetes, type 2 diabetes, diabetes unspecified and diabetes excluded.

### C. openEHR

At the core of openEHR is a generic reference model (RM) and specific *archetypes*, defined as constraint-based models of domain entities. The RM defines the structure of 1) archetype sets (COMPOSITION, SECTION); 2) archetype attributes (*datapoints*) (CLUSTER, ELEMENT, TREE); and 3) the specification of archetypes describing a specific clinical observation or interaction entry (EVALUATION, ACTION, OBSERVATION, INSTRUCTION, ADMINISTRATION ENTRY).

The openEHR Archetype Object Model (AOM) describes the definitive semantic model of archetypes in the form of an object model. It defines data types, constraints, and a reference mechanism allowing one archetype to reference another (*slot*). An archetype is essentially a set of *datapoints* defining characteristics of particular a clinical entry and is created under a specific reference model which mandates the types of information it can encapsulate. Archetypes are usually implemented via templates, a third layer overlaying the archetypes which allows the specification of additional restrictions (e.g. *cardinalities*, *optionality*) on the datapoints of one or more connected archetypes.

openEHR includes specifications of two semantic languages for archetype manipulation (Archetype Definition Language, ADL) and querying (Archetype Query Language, AQL). ADL is a formal language for expressing and serializing archetypes while AQL is a declarative query language specifically developed for searching clinical data found in archetype-based EHRs (no formal implementation exists).

## D. Desiderata for computable representations

We evaluated the ability of openEHR for providing computable representations of EHR phenotyping algorithms using the desiderata defined by Mo and colleagues [7]. In their work, the authors reviewed a series of EHR phenotyping algorithms which were developed as part of the Electronic Medical Records and Genomics (eMERGE) consortium [8], a national consortium of U.S. medical research institutions that combine DNA repositories with hospital EHR data in approximately 55,000 patients. Additionally, they reviewed a series of authoring tools for common features such as the Measure Authoring Tool (MAT) (https://www.emeasuretool.cms.gov/), i2b2 (https://www.i2b2.org/), and the SHARPn PhenotypePortal (http://phenotypeportal.org/) in order to extract common features.

The authors propose a list of recommendations for desired features (*desiderata*) for computable phenotype representation models:
- Support human-readable and computable representations
- Implement set operations and relational algebra
- Represent phenotype criteria using structured rules (e.g. nested logical structure, Boolean logic, comparative operations, aggregative operations, and negation)
- Support defining temporal relations between clinical events
- Utilize standardized controlled clinical terminologies and facilitate reuse value set reuse
- Support Natural Language Processing (NLP) operations for extracting clinically significant markers from unstructured EHR
- Provide interfaces for external software algorithms or data components
- Maintain backwards compatibility to accommodate for temporal changes in the underlying healthcare process model and EHR data specifications.

## III. RESULTS

### A. Archetype creation

No suitable diabetes archetypes were found in the public Clinical Knowledge Manager (CKM) archetype repository. As part of this work, we created a set of new archetypes for defining a diabetes diagnosis.

Created archetypes were organized into sections within a single *composition* (Fig. 2). Each archetype represents one phenotype component (e.g. diagnoses of diabetes in primary care) and *datapoints* represent the categories within components definition (e.g. the list of diagnostic terms associated with that component). *Datapoints* carry enumerations of diagnostic terms, which are bound to external terminologies (Read, ICD-9 and ICD-10). A template specifying a particular subject subclass (e.g. subject with type 1 diabetes) is then determined by adding additional constraints and cardinalities within the template with one template defining one subject subgroup.

openEHR phenotype components could be grouped based on their logical meaning into Symptoms, Diagnoses and Treatment or Procedure groups. In parallel to the above groupings, the EVALUATION, ACTION and OBSERVATION RM elements provide logical concepts which phenotype components/concepts could be expressed with (e.g. observed, objective symptoms could be expressed by an archetype inherited from the EVALUATION RM). Finally, RM structural concepts (COMPOSITION and SECTION) combined with the archetype *slot* mechanism provide a very limited way to express the actual phenotype algorithm logic.

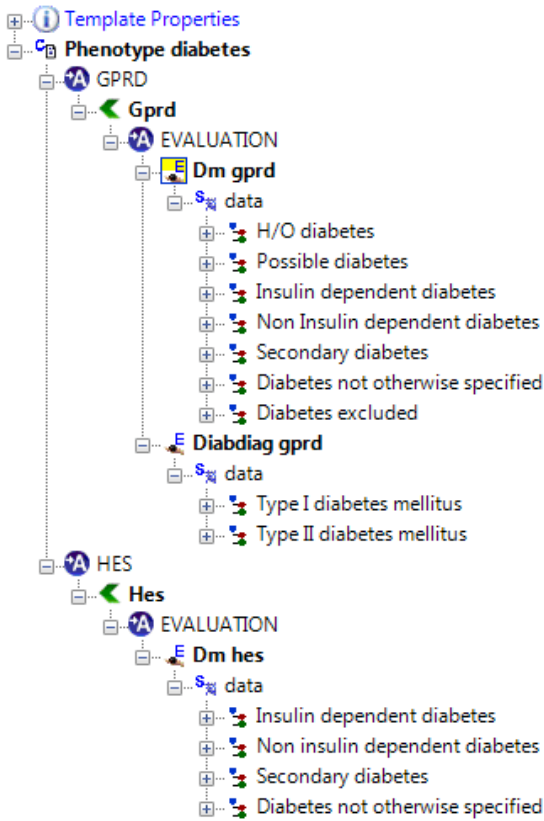

Figure 2. The created diabetes openEHR concept: individual phenotype components are represented as archetypes and contain lists of terms from controlled clinical terminologies.

In the following sections we evaluate the archetype driven diabetes EHR phenotyping algorithm representation against a series of desired characteristics (outlined in section D above) and discuss our findings.

B. *Human readable and computable representations*

openEHR archetypes support both human-readable and computable representations of phenotyping algorithms. Individual archetypes are machine-readable and conform to a predefined specification. Archetypes are defined in ADL and individual templates are stored as XML. An XML schema definition (XSD) is available to enable automated template validation and enable the transformation of archetypes into human-readable formats such as Hypertext Markup Language (HTML).

C. *Relational algebra operations*

openEHR does not natively support relational algebra operations and therefore phenotype logic operations cannot be fully expressed. A limited set of constructions substituting set operations do however exist. Archetypes can be considered as classes of instances with a set of specific attributes. Structural constructions (*sections*, *compositions*) allow us to group multiple archetypes, i.e. make a union operation above multiple classes and their respective instances. The additional constrains which are supported by templates could potentially act as a substitute for an exclusion operation (i.e. zero *datapoint* occurrence) and intersection (i.e. mandatory occurrence of *datapoints* from different archetypes). For example, a template representing a type 1 diabetes diagnosis derived from primary care EHR specifies zero occurrences of type 2 diabetes clinical terms and at least one clinical term related to type 1 diabetes. It is important to mention that templates were originally proposed with the EHR input forms as their main use-case and therefore set operations involving multiple archetypes are, by definition, limited.

D. *Structured and temporal phenotype criteria*

openEHR supports structured rules for defining phenotype criteria in a limited manner. The logical structure of the diabetes phenotype could be expressed by the RM structural models (*composition*, *section*) or by a set of tree-like structured datapoints (*clusters*). Multiple archetypes can refer to each other via *slots* enabling the creation of more complex, nested hierarchical structures. The diabetes phenotype archetype structure is composed of three *evaluation* archetypes (carrying clinical term enumerations) placed in two *section* archetypes (representing care settings, primary and secondary care), both connected to the main *composition* archetype.

openEHR supports all common data types (e.g. string/text, Boolean, integer/quantity, date-time), however comparative operations have to be implemented externally as they are not directly supported i.e. where a diagnosis is within 90 days of some other clinical event. Equally, aggregation and negation operations are not natively supported by and have to be handled by the upstream layer of the system implementation.

With regards to expressing temporal criteria, archetypes support definition of events as points in time as well as time intervals. Explicit definition of temporal relations however have to be handled by the upstream layer and are not directly supported.

### E. Standardized, re-usable nomenclature

Archetypes support a binding mechanism to pair internal values with external standardized controlled clinical terminology or classification system (e.g. ICD-9, ICD-10, Read, SNOMED-CT). All of the diagnostic terms used in the phenotyping algorithm could therefore be mapped to existing external ontological resources. Additionally, the Clinical Knowledge Management repository (www.openehr.org/ckm/) is an open-access catalogue of archetypes that supports their sharing of archetypes with collaborative reviewing.

### F. External data and interfacing

External interfacing was not directly tested in our work as the phenotyping algorithm used as a test case does not rely on external data (e.g. statistical distributions) nor does it require the integration of additional elements from external processes (e.g. natural language processing output). Since openEHR is designed primarily to define structured machine-readable EHR, NLP is not natively supported. Interfacing with external data and libraries can potentially be implemented using the openEHR Java API but would require significant effort.

### G. Backwards compatibility

openEHR is backwards compatible to the older openEHR EHRs. (ADL v2 to v1.4). Backward compatibility with older versions of clinical terminology systems is ensured by the archetype term binding system. Finally, since openEHR archetypes are essentially XML, this enables their storage within a revision control system (e.g. git) which can facilitate the tracking of changes during their development and evaluation in parallel.

## IV. DISCUSSION

In this work, we evaluated openEHR for storing EHR phenotyping algorithms in a machine-readable format (Table 1). openEHR is primarily designed to specify a structure of medical records with constrained-based models. However, since archetypes define the characteristics of a set of records and phenotyping algorithms are primarily based on a identifying patients fulfilling particular criteria (e.g. that have a diagnosis), archetypes could be used for defining explicit machine-readable phenotype definitions.

Phenotype representations defined in openEHR however do not satisfy all the desiderata proposed by Mo and colleagues and researchers are thus faced with significant challenges. NLP and external interface support is not natively handled by openEHR and has to be implemented by the upstream layer or though the Java API which would require significant amount of effort and additional code. Support for relational algebra, temporal criteria and structured rules is limited and that in turn limits the types of EHR phenotypes than can be defined. While the diabetes phenotype tested in this work does not rely on temporal restrictions, more complex disease phenotypes for non-chronic or acute manifestations would require these types of rules to be implemented.

TABLE I. THE ABILITY OF OPENEHR TO MEET THE DESIDERATA FOR COMPUTBABLE REPRESENTATIONS OF EHR PHENOTYPES

| Desiderata | Evaluation |
|---|---|
| Human-readable and computable representation | Fully supported – computable representations are based on XML and a supplied XSD schema can be shared via the Clinical Knowledge Management repository |
| Set operations and relational algebra | Not supported – these need to be implemented by the openEHR clinical information system |
| Structured rules | Not supported – these need to be implemented by the EHR information system |
| Temporal relations | Partially supported – archetypes support events as points in time or intervals but implementation relies on EHR information system |
| Standardized nomenclature | Fully supported – openEHR binding mechanism supports the pairing of internal values to external controlled clinical terminology systems |
| External interfacing | Partially supported – openEHR provides a Java-based API for interfacing with external software algorithms, Natural Language Processing systems or other interfaces |
| Backward compatibility | Fully supported – XML documents for archetypes can be additionally stored in revision control systems to track changes over time |

Using openEHR as a format for storing computable representations of EHR phenotypes does not seem to be the ideal choice and further work would be required to transform the standard to fully support the required operations.

ACKNOWLEDGMENT

This study was supported by the National Institute for Health Research (RP-PG-0407–10314), Wellcome Trust (086091/Z/08/Z), the Medical Research Council Prognosis Research Strategy Partnership (G0902393/99558) and the Farr Institute of Health Informatics Research, funded by The Medical Research Council (K006584/1), in partnership with Arthritis Research UK, the British Heart Foundation, Cancer Research UK, the Economic and Social Research Council, the Engineering and Physical Sciences Research Council, the National Institute of Health Research, the National Institute for Social Care and Health Research (Welsh Assembly Government), the Chief Scientist Office (Scottish Government Health Directorates) and the Wellcome Trust.